\documentstyle[preprint,aps,epsf,eqsecnum,psfig,epsfig,tighten]{revtex} 
\preprint{\vbox{\baselineskip=12pt 
\rightline{CGPG-96/3-8} 
}} 

\def\be{\nopagebreak[3]\begin{equation}} 
\def\ee{\end{equation}} 
\def\ba{\nopagebreak[3]\begin{eqnarray}} 
\def\ea{\end{eqnarray}}

\def\c{\gamma} 
\def\d{\delta} 
\def\e{\eta} 
 
\def\f{\phi}

\def\m{\mu}

\def\p{\pi}

\def\s{\sigma} 
\def\t{\tau}

\def\S{\Sigma}

\newcommand{\teta}{\rlap{\lower2ex\hbox{$\,\tilde{}$}}\eta{}}
\newcommand{\tiN}{\rlap{\lower2ex\hbox{$\,\tilde{}$}}N{}}

\newcommand{\tE}{\tilde{E}}

\begin{document}
\draft
\title{
Combinatorial space from loop quantum gravity. 
}
\author {
Jos\'e A. Zapata
\thanks{zapata@phys.psu.edu}
\thanks{
Department of Physics, 
The Pennsylvania State University, 
104 Davey Laboratory, University Park, PA 16802}
}
\address{}

\maketitle
\begin{abstract} 
The canonical quantization of diffeomorphism invariant 
theories of connections in terms of loop variables is revisited. 
Such theories include general relativity described in terms of 
Ashtekar-Barbero variables and extension to 
Yang-Mills fields (with or without fermions) coupled to gravity. 

It is argued that the operators induced by classical 
diffeomorphism invariant or covariant functions are 
respectably invariant or covariant under 
a suitable completion of the diffeomorphism group. 
The canonical quantization in terms of loop variables 
described here, yields a representation of the algebra of
observables in a {\em separable} Hilbert space. Furthermore, the 
resulting quantum theory is equivalent to a model 
for diffeomorphism invariant gauge theories which replaces 
space with a manifestly combinatorial object. 
\end{abstract}
\clearpage 



\section{Introduction}

After ten years of `new variables' \cite{ashtekar} 
and loop representation \cite{loop}, the theory 
has matured significantly. This approach to quantum gravity 
has gained clarity, borrowed and developed powerful tools, 
and sharpened its picture of physical space.  
Specifically, after solving the spin (Mandelstam) identities 
by the use of spin networks \cite{s-n} 
the formulation of the theory 
has become clear and it allows a deeper understanding.  
After this clarification was made, 
explicit geometric operators \cite{geoperators} that 
encode loop quantum gravity's picture of space were written. 
These geometric operators predict a geometry that is 
polymer-like \cite{polymer}, non-commutative \cite{non-comm} and 
quantized \cite{geoperators}. 
Also lattice versions of the framework \cite{lattice,c} 
ready for explicit computation, and several proposals for the 
Hamiltonian constraint of the theory \cite{ham,thomasham} 
have been developed. 
Now the early results (on the classical/macroscopic limit 
\cite{weeves} and incorporating other fields and matter 
\cite{carlo-hugo-kiril}) have to be ``upgraded,'' 
and using the new tools and sharper notions other problems, 
like the computation of transition amplitudes \cite{trans-ampl} or the 
statistical mechanics behind black hole thermodynamics \cite{bh}, 
are within reach. 


Loop quantization 
\cite{analytic,loopdpr} applies to any gauge theory with compact 
gauge group and particularly to general relativity 
casted in the Ashtekar-Barbero variables \cite{barbero}. 
Information about the connection and the triad 
is stored in a set of functions of the holonomies along 
paths and a set of functions of the surface smeared triads 
respectively. Loop quantization produces an operator for every 
element of this family of functions. 
This is done by completing the space of holonomy 
functions to make it a $C^*$ algebra; 
and turning it into a Hilbert space by 
giving it an inner product 
that makes the operators induced by former real valued functions 
be Hermitian operators. 

To represent the algebra of observables one needs a 
space of invariant states. Since quantization involves 
completing the algebra of holonomy functions, the quantum gauge 
group is an appropriate completion of the classical gauge group. 
In the case of the internal gauge 
transformations one can solve the Gauss constraint after quantization, 
or give an 
intrinsically gauge invariant formulation. Both constructions agree 
if the quantum gauge group is taken to be a completion of the 
classical internal gauge group. 
For the diffeomorphism gauge group there is no intrinsically invariant 
construction; one can only solve the diffeomorphism constraint after 
quantization. In this article I argue that there is a natural 
candidate for the quantum gauge group, and it turns out to be a 
completion of the diffeomorphism group. 

According to this refined treatment of diffeomorphism 
invariance an old expectation is realized. Namely, diffeomorphism 
invariance plays a double role. It forces one to 
consider an uncountable set of graphs to label the kinematical 
states of loop quantum gravity. However, it 
yields a representation of the algebra of observables
(diffeomorphism invariant functions) in a 
{\em separable} Hilbert space spanned by states labeled by 
knot-classes of graphs. 
In contrast, Grot and Rovelli found that the space of invariant 
states of the previous formulation of loop quantization 
contains families of orthogonal states 
labeled by continuous parameters \cite{non-sepa}. 
In the version of loop quantization 
that uses the completed diffeomorphism group, 
one can exhibit a 
countable basis of invariant states (the spin-knot basis). 
In fact, the completed diffeomorphism group simplifies 
the formalism, and the resulting 
quantum theory is equivalent to a model 
for diffeomorphism invariant gauge theories which replaces 
the space manifold with a manifestly combinatorial object \cite{c}.
Just as loop quantization conduces to a notion of quantum geometry with 
discrete areas of non-commutative nature, it also conduces to an 
intrinsically combinatorial picture of physical space. 

I revisit loop quantization 
emphasizing the issue of diffeomorphism invariance. 
For completeness, 
the kinematics of loop quantization is briefly reviewed in 
section~\ref{2}. 
Internal and diffeomorphism 
gauge invariance of the classical and quantum 
theories are addressed in section~\ref{3}--the main section of 
the article. In that section, a refined 
treatment of diffeomorphism invariance is 
presented, and its consequences are studied. A discussion section 
ends the article.

\section{Kinematics: 
general relativity in terms of connections and holonomies}
\label{2}


Recall that gravity, expressed in (real) Ashtekar-Barbero variables, 
is a Hamiltonian theory of connections that shares the phase 
space with 
$SU(2)$ Yang-Mills theory \cite{barbero}. 
That is, the configuration variable is a connection 
$A_a^i$ taking values in the Lie algebra of $SU(2)$, and the 
canonically conjugate momentum is 
a triad $\tE^a_i$ of densitized vector fields. 
In these variables the contravariant spatial metric is 
determined by 
$q^{ab} \det{q}= \tE^a_i \tE^{bi}$, which makes contact with the 
usual geometrodynamic treatment of general relativity. 
In this formulation, Einstein's equations are equivalent to a 
series of
constraints: a set which generates diffeomorphisms on the Cauchy
surface 
and constitutes a closed subalgebra of the constraint algebra, 
and a set of constraints generating motions transverse to the initial 
data surface. 

If only the constraints that generate spatial 
diffeomorphisms are imposed and the Hamiltonian constraint is dropped, 
one gets a well-defined model to study diffeomorphism 
invariant theories of connections. This model is called the 
Husain-Kucha\v{r} model and can be derived from an action principle 
\cite{husain-kuchar}; it shares the phase space, the gauge constraint 
and the diffeomorphism constraint with general relativity and 
has local degrees of freedom. More than 
a toy model, the Husain-Kucha\v{r} model provides 
an intermediate step in the quantization of general relativity; 
a quantization of the model requires to set up 
a kinematical framework and regularize and solve 
the gauge and diffeomorphism constraints. After a satisfactory quantum 
version of the model is developed, a quantization of 
general relativity amounts to the difficult tasks of 
regularizing and solving the Hamiltonian constraint and verifying that
GR is 
recovered in the classical limit. This article is about 
the treatment of diffeomorphism invariance in the loop quantization 
framework; therefore it pertains to any 
diffeomorphism invariant theory of connections, in 
particular, to general relativity (possibly coupled to Yang-Mills
fields) 
and the Husain-Kucha\v{r} model. For the sake of 
concreteness, the problems and the results are stated in reference  
to the the quantization of the Husain-Kucha\v{r} model. 
Issues like whether the algebra of the constraints is correct or if
there 
is a classical limit in the theory resulting from Thiemann's
Hamiltonian 
constraint \cite{thomasham} is matter of hot debate \cite{wrong}. 
Since the study includes diffeomorphism covariant functions 
and the Hamiltonian constraint is diffeomorphism covariant, 
the results presented in this article may shine some 
light on the difficult problem of regularizing the Hamiltonian
constraint.


The cornerstone of loop quantization is the use of holonomies 
along loops as ``coordinates on the classical configuration space'' 
\cite{loop,loopdpr}. 
For primary momentum functions one can use 
the triad (whose dual is a two form) smeared on surfaces 
\cite{geoperators,non-comm}, 
or, in the manifestly gauge invariant treatment, a combination 
of holonomies and triads called the strip functions
\cite{strip,non-comm}. 
In this article, the term 
{\em loop variables} is some times used as a collective 
name for the configuration and momentum variables described. 
This choice of variables is due to 
the symmetries of the theory; using them one can explicitly 
solve the gauge and diffeomorphism constraints of the quantum 
theory. 

It was proven \cite{giles-lewandowski} that all the 
information about the connection is contained in the set of 
holonomies of the connection around every smooth path $e$ 
\be
h_e(A)= \rm{Pexp}(i\int_e \t_i A_a^ids^a)
\ee 
where $\t_i=\frac{1}{2}\s_i$ 
are the $SU(2)$ generators \cite{loopdpr}. 
The loop variables $h_e(A)$ are an 
overcomplete set of configuration functions that coordinatize the 
space of smooth connections ${\cal A}$ in the sense that two 
connections can always be differentiated by the loop variables. 
If only closed loops are used, the set of traces of the holonomies 
coordinatizes the space of 
connections modulo internal gauge transformations. 
Also, any two smooth triads can be differentiated by smearing the
triads 
(two forms) over some surface. 
This property ensures that by keeping only functions of the loop
variables 
as primary functions, and recovering 
every thing from them after quantization, 
no relevant information is omitted. 
Thus, at least in principle, any phase space function can be 
expressed in terms of functions of the loop variables. 
The holonomy functions are special because they form a subalgebra 
of the algebra of configuration functions; and this subalgebra 
is preserved 
by the primary momentum functions, the surface smeared triads. 
These important properties lie at the heart of loop quantization. 

The classical algebra that is actually quantized 
is the algebra ${\rm Cyl}_0$. 
A cylindrical function $f_{\c}(A) \in {\rm Cyl}_0$ is 
a function of the holonomies along the edges of the graph $\c$. With
this 
definition, the product of two cylindrical functions is another
cylindrical 
function if the edges of the two original graphs are contained in the
set 
of edges of a bigger graph. To satisfy this condition, it was first
proposed 
to consider only graphs with piecewise analytic edges \cite{analytic}. 
Since among the cylindrical functions one has all the loop variables,
it is 
clear that one can use the cylindrical functions as primary functions
in 
the space of smooth%
\footnote{
I would loosely use the term smooth to mean real analytic; except in
the 
last paragraphs of the article where I comment on the smooth 
($C^{\infty}$) category. 
} 
connections. 
After ${\rm Cyl}_0$ is quantized the primary 
configuration functions become 
operators that act by multiplication, and the primary momentum
functions 
(the surface smeared triads) become operators that act as derivative 
operators. Thus, loop quantization produces a regularized operator 
from any phase space function written in terms of 
the loop variables.

${\rm Cyl}_0$ is quantized by following a series of steps. First, 
completing it to form a $C^*$ algebra ${\rm Cyl}$. 
Second, represent the cylindrical functions and linear in 
momenta functions in 
${\cal H}_{\rm kin} = L^2(\bar{\cal A}, \m)$ 
(by multiplicative and derivative operators respectively), where 
$\bar{\cal A}$ is the spectrum of ${\rm Cyl}$ and $\m$ is the 
Ashtekar-Lewandowski measure, 
which is selected by the reality conditions \cite{analytic}. 

At a more operational level, the Hilbert space of 
gauge invariant states (under $SU(2)$ gauge transformations) 
is spanned by spin network states $|S\rangle$ \cite{s-n}. 
A spin network $S$ is labeled by a colored graph $\vec{\c}$ 
and represents the function of the holonomies 
along its edges given by 
\be \label{S} 
S_{\vec{\c} , j(e), c(v)}(A) = 
\prod_{e\in E_{\vec{\c}}} \p _{j(e)}[ h_e(A) ]\cdot 
\prod_{ v\in V_{ \vec{\c} } } c(v) \quad , 
\ee 
where the colors on the edges $j(e)$ are irreducible 
representations of $SU(2)$, and the vertices are labeled by 
gauge invariant contractors 
$c(v)$ that match all the indices 
(in the formula denoted by `$\cdot$') 
of the holonomies of the edges. An inner product in the space of gauge 
invariant states $L^2(\overline{\cal A/G}, \m)$ 
is given, alternatively, by the Ashtekar-Lewandowski measure 
\cite{almeasure,analytic} or by recoupling theory 
\cite{loopdpr,recoupling}. 
According to this inner product, two spin network states are orthogonal
if 
their coloring or labeling graphs are different. 
Using a convenient set of contractors one can form an 
orthonormal basis 
with spin network states \cite{recoupling} 
\be 
\langle S|S' \rangle = \d_{S S'} \quad . 
\ee 
Non-gauge invariant spin network states 
are constructed by just dropping the gauge invariant 
contractors and the Ashtekar-Lewandowski measure induces an inner 
product in ${\cal H}_{\rm kin}$. 

\section{Diffeomorphism invariance 
in the classical and quantum theories} 
\label{3}

Classical observables, 
gauge and diffeomorphism invariant functions, induce 
functions in the reduced phase space; loop quantization's 
objective is to produce a faithful representation of the algebra of 
observables. 
First the operators are regularized from their expressions 
as functions of the loop variables. The resulting operators 
are expected to be invariant under ``quantum gauge transformations'' 
and ``quantum diffeomorphisms.'' 
Finally, from the algebra of invariant operators one 
induces (by dual action) a faithful representation on the space 
of diffeomorphism invariant states. 
Here, this process is followed, but 
special care is paid to the character acquired by diffeomorphism 
invariance after loop quantization. 

In the description of the classical theory in terms of smooth fields
there 
is a harmony between the space of smooth connections and the gauge
group. 
As far as the internal gauge transformations, the internal gauge group
may be  
characterized as the set of $SU(2)$-matrix valued functions $g$ such
that 
given a smooth connection $A \in {\cal A}$, the connection 
$g(A_a)= g^{-1} A_a g + g^{-1} \partial_a g$ is also smooth. 
Similarly, the diffeomorphism group can be characterized as the
subgroup 
of the homeomorphism group composed by all the transformations which 
leave the space of smooth connections invariant 
\be \label{har}
{\rm Diff} = \{ \f \in Hom | \f^*(A) \in {\cal A} 
\quad \hbox{for all} \quad A \in {\cal A} \}
\ee

This compatibility between configuration space and gauge group 
acquires a different form after loop quantization. 
Quantization takes the space of smooth connections and, by 
completing it, constructs 
the quantum configuration space $\bar{\cal A}$. 
A generalized connection $A \in \bar{\cal A}$ 
simply assigns group elements to piecewise analytic paths; that is, it 
acts as a connection which does not need to be smooth. 
Completing the configuration space requires adapting the gauge group
also. 
The quantum internal gauge group $\bar{\cal G}$ is formed by the 
transformations acting at the end points of the paths, 
$g(A) [e] = g^{-1}(e_0) g(A) [e] g(e_1)$. 
A quantum gauge transformation maps 
every generalized connection to another generalized connection. This
group
contains the classical internal gauge group, but it is not the
classical gauge
group. It is the completion of the group of smooth internal gauge
transformations according to the operator norm. 
Most of the quantum gauge transformations would transform a smooth
connection
into a non-smooth connection.

In the diffeomorphism part of the gauge group a similar phenomena
happens. 
The family of piecewise analytic graphs is 
left invariant by a bigger group than the group of smooth
diffeomorphisms, 
but if one transforms a smooth connection using a non-smooth map one 
obtains a non-smooth connection. Again, because quantization 
involves completing the configuration space, the generalized 
connections are covariant with respect to a certain completion of 
the diffeomorphism group;  
$\f ^* (A) [e] := A [\f(e)]$ is defined for a certain completion of 
the diffeomorphism group%
\footnote{
In the previous paragraph I defined $\bar{\cal G}$ algebraically. 
The algebraic relation came from the classical theory, but the definition
of $\bar{\cal G}$ only involved quantum objects. I will show that 
this construction in the case of the diffeomorphism group yields 
$\bar{D}$. However, $\bar{\cal G}$ is the completion of ${\cal G}$ in the
operator norm, and $\bar{D} \supset D$, but according to the operator
norm $\bar{D}$ is a discrete group. Strictly speaking, 
$\bar{D}$ is an algebraic extension
of the diffeomorphism group rather than a completion of it. 
}. 
As a consequence, 
the primary configuration and momentum variables 
induce operators that are covariant with respect to the  
mentioned completion of the diffeomorphism group. 
Since every operator of the quantum theory is constructed from the 
primary configuration and momentum operators, 
this extended covariance becomes a feature of the quantum theory. 
Functions of the phase space 
with a geometrical label (like the holonomy functions, surface smeared 
triads, surface area functions, volume functions, etc) are 
diffeomorphism covariant, but operators coming from 
these functions with geometrical labels are 
naturally covariant under a 
certain completion of the diffeomorphism group. 
Note that the Hamiltonian constraint is diffeomorphism covariant and 
some of its regularizations have the mentioned extended covariance 
(comments on the Hamiltonian constraint are reserved for 
the discussion section). 

More importantly, given the extended notion of covariance, 
it is necessary to review the notion of observable in the quantum
theory. 
Observables (diffeomorphism invariant functions) 
naturally arise from covariant functions where the geometrical labels 
become dynamical. For example, 
area functions of surfaces specified by matter 
fields. 
If the fields specifying the geometrical labels also acquire the 
extended covariance, as they would if they are quantized using loop 
quantization, then the natural notion of an observable would be 
to be invariant under the mentioned completion of the 
diffeomorphism group. 

To explain the details of the previous discussion let me show you how 
piecewise analytic diffeomorphisms come about. 
Consider the following situation. The Cauchy 
surface is $R^3$; an example of nonsmooth map is 
$\f:R^3 \to R^3$ defined to be 
the identity above the $x-y$ plane and below the 
plane $x-y$ plane it is defined by $\f (x,y,z) = (x, y+mz, z)$. 
This map is smooth above and below the $x-y$ plane but at the $x-y$
plane 
its derivative from above and its 
derivative from bellow do not match (in the direction normal to the 
$x-y$ plane). One can see that $\f$ maps some 
smooth loops to loops with kinks. 
Given any smooth connection 
$A \in {\cal A}$, 
one would like to say that the functions 
\be 
h_l(\f ^* (A)) := h_{\f (l)}(A) \quad .
\ee 
are ``covariantly'' related to the loop coordinates of $A \in {\cal
A}$, 
but the connection $A^\prime =\f ^* (A)$ is not in the configuration 
space of the classical theory. However, 
in the quantum theory, the functions 
$h_{\f (l)}(A)$ induce an operator that is as valid as the 
ones induced by the functions 
$h_{\pi (l)}(A)$ defined using any smooth map $\pi$. 
Hence the map $\f$ is an object that will play a role in the quantum
theory 
even though it did not define a canonical transformation in the 
classical theory. 
Classically, we cannot ask if the connections $A \in {\cal A}$ and 
$A^\prime =\f ^* (A)$ are gauge related, but the quantum configuration 
space is the space of generalized connections, 
and $A \in \bar{\cal A}$ if and only if $\f ^* (A) \in \bar{\cal A}$.

Following the above example, 
a map $\f :\S \to \S$, that maps any piecewise analytic graph to 
another, would map any generalized connection to another, and 
define a new loop operator from a given one. 

{\em A map $\f :\S \to \S$ belongs to $\bar{D}$ iff 
for any piecewise analytic graph $\c$ the new graph $\f(\c)$ is also 
piecewise analytic. }

Above I gave a description of $\bar{D}$ designed to show 
the natural role that it will play in the quantum theory, 
and to emphasize the parallelism between its definition and the 
definition of $\bar{\cal G}$. 
Alternatively, one can describe 
$\bar{D}$ as the group of piecewise analytic diffeomorphisms. 
In close analogy with the definition of a piecewise linear manifold 
(Regge lattice), a 
piecewise analytic manifold $\S$ is a topological manifold formed as a 
union of finitely many 
closed cells, each of which is an analytic 
manifold with boundary (these correspond to the higher dimensional 
simplices of the Regge lattice). 
Two of these cells may intersect only at their boundaries. 
A map $\f :\S_1 \to \S_2$ is piecewise analytic if and 
only if there is a refinement of the cell decomposition of 
$\S_1$ such that the restriction of 
$\f$ to every cell is an analytic map. 
Clear examples of piecewise analytic manifolds (maps) 
are real-analytic manifolds (maps) and piecewise linear manifolds
(maps).

Guidance from the classical theory tells us that the operators induced
by 
$h_l(A)$ and $h_{\pi (l)}(A)$ for any smooth map $\pi$ are gauge
related. 
However, classically one can not say that the functions 
$h_l(A)$ and $h_l(\f ^* (A)) := h_{\f (l)}(A)$ are gauge related since 
the non-smooth map $\f$ does not define a 
canonical transformation because 
the connection $A^\prime =\f ^* (A)$ is not in the configuration 
space of the classical theory, 
but the quantum states are functions of generalized connections
${\rm Cyl}(\bar{\cal A})$ 
and $A \in \bar{\cal A}$ if and only if $\f ^* (A) \in \bar{\cal A}$. 
Quantization involves completing the space of cylindrical functions to
make 
it the $C^*$ algebra ${\rm Cyl}(\bar{\cal A})$; to account for this 
enlargement of the configuration space, the internal gauge group is 
$\bar{\cal G}$ instead of ${\cal G}$. 
Smooth connections and generalized connections differ in more than
their 
``internal degrees of freedom.'' Recall that in the smooth case 
$\f ^* (A)$ is defined only for smooth (analytic) maps, whereas in 
the case of generalized connections it is defined for any piecewise 
analytic map.

Because of these considerations, and 
since any piecewise analytic map $\f$ 
can be obtained as a limit of smooth maps 
I will assume that the operators induced by $h_l(A)$ and $h_{\f
(l)}(A)$ 
are gauge related.

A quantum `diffeomorphism' $\f \in \bar{D}$
acts by shifting the labels of the spin networks by a diffeomorphism 
\be 
U_\f |
S_{\vec{\c} , j(e'), c(v')} \rangle := 
| S_{\vec{\f(\c)} , j(e'), c(v')} \rangle \quad . 
\ee 
Since the measure that defines the inner product is $\bar{D}$ 
invariant, the operator $U_\f$ is unitary. 

Before the significance of $\bar{D}$ 
was understood, it was noticed that the original regularization of the 
area and volume operators, and some versions of the Hamiltonian
constraint, 
were not diffeomorphism covariant, but they were covariant under a 
bigger group. Later a version of the volume operator that was only
covariant 
under smooth diffeomorphisms 
was developed and this version of the volume operator
entered in the definition of Thieman's Hamiltonian constraint.
Initially, 
it was believed that replacing 
the volume operator used by Thieman with the $\bar{D}$ covariant 
version would change the algebra of the constraints, but now it 
has been proven that it produces no changes \cite{wrong}.

Using the technique developed in \cite{analytic}, one solves the 
quantum diffeomorphism constraint by 
constructing the space of $\bar{D}$ invariant 
states ${\cal H}_{\rm diff}$. It is spanned by 
s-knot states $\langle s|$, labeled by knot-classes of colored graphs, 
and defined by 
\be \label{s}
\langle s_{[\vec{\c}] , j(e), c(v)}| 
S^{\prime}_{\vec{\e} , j(e'), c(v')} \rangle 
:= a([\c]) \d_{[\c] [\e]} \sum_{[\f]\in {\rm GS}(\c)} 
\langle S_{\vec{\c} , j(e), c(v)}| U_{f\cdot \f_0}
S^{\prime}_{\vec{\e} , j(e'), c(v')} \rangle  
\ee 
where $a([\c])$ is an undetermined normalization parameter, 
$\d_{[\c] [\e]}$ is non vanishing only if there is 
a {\em piecewise analytic 
diffeomorphism} $\f_0 \in \bar{D}$ that maps $\e$ to a graph 
$\c$ that defines the knot-class $[\c]$, 
and $\f \in \bar{D}$ is any element in the class of 
$[\f]\in {\rm GS}(\c)$. The finite group ${\rm GS}(\c)$ is the group of 
symmetries of $\c$; in other words, 
the elements of ${\rm GS}(\c)$ are maps between the 
edges of $\c$ (for a detailed explanation see \cite{analytic,c}). 

The s-knot states are solutions of the diffeomorphism constraint 
because its action is invariant under quantum diffeomorphisms 
by construction. 
An inner product for ${\cal H}_{\rm diff}$ 
is given simply by%
\footnote{
Note that this inner product is determined only up to 
the unknown parameters $a([\c])$. 
} \cite{analytic} 
\be 
\langle s_{[\vec{\c}] , j(e), c(v)}| 
s^{\prime}_{[\vec{\d}] , j(e'), c(v')} \rangle 
:= \langle s^{\prime}_{[\vec{\d}] , j(e'), c(v')} | 
S_{\vec{\c} , j(e), c(v)}\rangle 
\ee 

The observables of the Husain-Kucha\v{r} model are naturally 
represented on ${\cal H}_{\rm diff}$. If 
$\hat{O}$ is a ``diffeomorphism'' invariant Hermitian 
operator on the kinematical Hilbert space, 
$\tilde{O}: {\cal H}_{\rm diff} \to 
{\cal H}_{\rm diff}$ is defined by its dual action 
\be 
(\langle s_{[\vec{\c}] , j(e), c(v)}| \tilde{O}) 
|S_{\vec{\c} , j(e), c(v)}\rangle := 
\langle s_{[\vec{\c}] , j(e), c(v)}| 
(\tilde{O} |S_{\vec{\c} , j(e), c(v)}\rangle ) \quad .
\ee 

These are the foundations of the theory 
following from considering the extended notion of diffeomorphism 
covariance/invariance in loop quantization.  
In particular, they constitute a quantization of the 
Husain-Kucha\v{r} model \cite{husain-kuchar}, 
that has local degrees of freedom. 

Here I will describe the properties of the quantum theory that 
are not shared by previous treatments of loop quantization. 
First, one should notice that ${\cal H}_{\rm diff}$ is separable. 
The s-knot states are labeled 
by knot-classes of graphs $[\c]$ with respect to 
$\bar{D}$. Since the diffeomorphism group was replaced by a 
bigger group, the resulting knot-classes are 
much bigger and therefore there are very few of them; this is why 
separability arises. In contrast, 
states in the original treatment are labeled by continuous parameters 
parameterizing the knot-classes of graphs with higher valence vertices 
\cite{non-sepa}. 

I sketch the proof of separability in the next few paragraphs. 
A mathematically rigorous proof can be found in the appendix of
\cite{c}.

Consider a three dimensional 
triangulated manifold $|K|$, which can be thought of as a three
dimensional 
Regge lattice. Since the interior of the tetrahedrons of the 
lattice are flat, one can define the baricenter of 
any simplex (tetrahedron, face or link); by adding these points 
to the original lattice, and also adding new links and faces (see fig.
1), 
one constructs the finer lattice $|Sd(K)|$ called the 
{\em baricentric subdivision} of 
the original lattice $|K|$. One can do this subdivision again 
and again to get a sequence of lattices 
$\{ |K|, |Sd(K)|, \ldots , |Sd^n(K)|, \ldots \}$. 
All these lattices are not disconnected, 
they are all subdivisions of $|K|$; 
using them, one defines a 
combinatorial graph $\c_c$ to be a graph in $|K|$ all whose edges are 
links of some of the refined lattices $|Sd^n(K)|$. 
Also consider a {\em fixed} map $h : |K| \to \S$ that maps every 
combinatorial graph $\c_c$ to a piecewise analytic 
graph $h (\c_c)$ on $\S$. 

\nopagebreak
\hskip .5in \psfig{figure=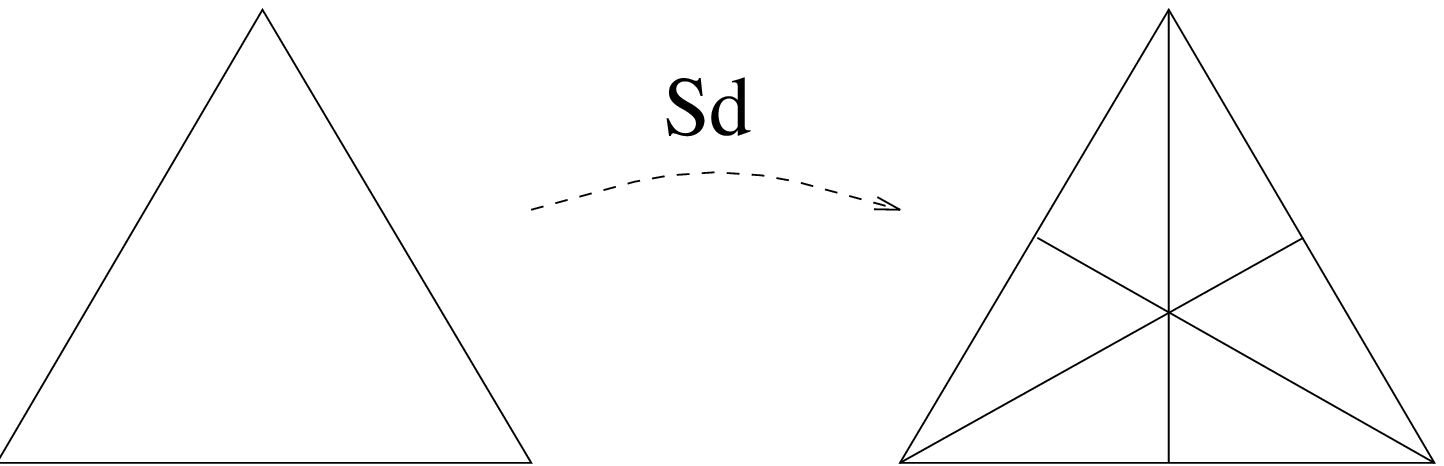,height=1.5in}

{\small {\bf Fig. 1}
A triangular face and its baricentric subdivision. Every link of $|K|$ 
is divided into two links of $|Sd(K)|$, every face into six faces 
and every cell into twenty four cells of $|Sd(K)|$. 
{
\label{figpa}

The sense in which the knot-classes of graphs $[\c]$ are big is that 
every class contains a combinatorial graph, $h (\c_c) \in [\c]$. 
Given an arbitrary graph $\c$, the following series of steps 
generates a combinatorial graph $\c_c$ and a piecewise analytic map 
$\f : \S \to \S$ such that $\f (h (\c_c)) = \c$. 
\begin{enumerate} 
\item \label{11}
Find $n$ 
such that $|Sd^n(K)|$ separates the vertices of 
$h^{-1}(\c)$ to 
lie in different simplices. (The conventions are such that 
every point of the manifold belongs to 
the interior of one and only one simplex of a given triangulation). 

\item \label{12}
Let $h_1:|K| \to |K|$ be the piecewise linear 
map that fixes the vertices of 
$|Sd^n(K)|$ and sends the new vertices 
$v \in |Sd^{n+1}(K)|$ (the baricenters of the simplices of 
$|Sd^n(K)|$) to: 
\begin{enumerate} 
\item themselves ($h(v)=v$), if there is no vertex of 
$h^{-1}(\c)$ in the simplex of $|Sd^n(K)|$ which has $v$ as baricenter.

\item the vertex of the graph ($h(v)= w$), in the case when 
the simplex of $|Sd^n(K)|$ which has $v$ as baricenter contains a 
vertex of the graph ($w \in h^{-1}(\c)$) in its interior. 
\end{enumerate} 

\item \label{13}
Find $m$ such that 
$h_1(|Sd^{n+m}(K)|)$ separates the edges of $h^{-1}(\c)$ 
in the interiors of different simplices%
\footnote{
In the case of a graph $\c$ with two or more edges meeting at a vertex
this step needs to be refined. One needs to find an integer $m$ and a 
piecewise analytic map $\psi :\S \to \S$ (with analycity domains given by
$h \circ h_1(|Sd^{n+m}(K)|)$ see next footnote) such that 
$\psi \circ h \circ h_1(|Sd^{n+m}(K)|)$ separates the edges of $\c$. 
Using this refinement, the rest of the construction has a clear 
extension. 
}. 

\item \label{14}
Let a {\em cell} be a (closed) image (by $h : |K| \to \S$) 
of a simplex of $h_1(|Sd^{n+m}(K)|)$. 
Let $\f=\f_1 \circ h \circ h_1 \circ h^{-1}:\S \to \S$, 
where $\f_1$ is a piecewise analytic map that is equal to the identity 
when restricted to cells which do not intersect $\c$, and 
sends the cells which intersect $\c$ to themselves, 
but has nontrivial analycity domains%
\footnote{
A piecewise analytic map is a continuous 
map whose restriction to the interior of 
any of its analycity domains is analytic. 
}.  
The analycity domains divide the cell 
into the {\em subcells} 
given by the image (by $h : |K| \to \S$) of 
the simplices of $h_1(|Sd^{n+m+1}(K)|)$. 
$\f_1$ must be such that the intersection of $\c$ and the cell lies in 
the the image (by $\f_1$) of the boundaries of the subcells; since 
only one (analytic) edge of $\c$ intersects the 
interior of the original cell, a map $\f_1$ 
with the requested property always exists. 
\end{enumerate} 
From the construction of $\f : \S \to \S$ it is immediate that 
$\f (h (\c_c)) = \c$.

The sense in which there are very few knot-classes of graphs 
is that the set of combinatorial graphs 
$\{ \c_c \}$ is countable. One can easily convince oneself 
that this is 
the case because every $\c_c$ belongs to $|Sd^n(K)|$ for some $n$, and 
there are countably many of these triangulations, each of which has
finitely 
many links%
\footnote{
One can triangulate a compact manifold with finitely many simplices and
a 
paracompact manifold 
with countably many simplices. I sketch in the argument for the 
compact case, but it is immediate to extend it 
to the paracompact case, 
which includes all the Cauchy surfaces of asymptotically flat
spacetimes. 
}. 
This property implies that the set of labels of 
the s-knot states is countable; that is, the Hilbert space of 
`diffeomorphism' invariant states ${\cal H}_{\rm diff}$ 
is separable.

I used the combinatorial graphs to prove the separability of the 
Hilbert space, but there is a deeper consequence of 
the existence of such a subspace of ${\cal H}_{\rm kin}$.  
It has a manifestly combinatorial origin and is capable 
of generating all the states in the space of solutions to the 
diffeomorphism constraint. As far as observables are concerned, the 
combinatorial states are sufficient; meaning that the 
manifestly combinatorial framework yields a unitarily 
equivalent representation of the algebra of observables 
(see the appendix of \cite{c} for a rigorous proof). 

Equivalence with a manifestly combinatorial model is not so surprising 
if one remembers that observables in generally covariant theories are 
supposed to measure only relative `positions' of the dynamical fields. 
One may object that in pure gravity there are not enough explicitly 
known observables as to serve as a basis of any argument. 
But, physically meaningful observables will 
arise if other fields are coupled to 
pure gravity (or to the Husain-Kucha\v{r} model). 
In these systems one can study observables that measure the 
gravitational field; for example, any covariant operator of pure
gravity, 
say an area operator, 
whose labeling 
surface becomes dynamical after coupling other fields becomes an 
observable. They are generally covariant systems with plenty of
observables 
measuring the gravitational field. 
Proving equivalence with a manifestly combinatorial model 
explicitly exhibits the relational nature of loop quantization.

In contrast with the treatment of diffeomorphism 
invariance presented in 
this article, the original study of the quantization of the 
Husain-Kucha\v{r} model considered the diffeomorphism group 
as the quantum gauge group. By using the same kinematical 
Hilbert space, but averaging over the diffeomorphism group instead of 
$\bar{D}$ to generate the solutions of the diffeomorphism 
constraint (\ref{s}), they constructed the space of ``physical'' states 
${\cal H}_{\rm Diff^*}$. 
This difference implies, 
in particular, that ${\cal H}_{\rm Diff^*}$ is not separable 
\cite{non-sepa} and that the nature of the theory is not combinatorial.

It was argued that classical functions which are diffeomorphism 
invariant/covariant induce, after loop regularization, 
$\bar{D}$ invariant/covariant 
operators on ${\cal H}_{\rm kin}$. Because the operators are invariant 
under a larger group, 
the algebra of observables can be represented in ${\cal H}_{\rm
Diff^*}$; 
however, the representation of such operators yields a continuum of 
superselected sectors \cite{thetasuperselction}. This superselection 
is not surprising after one knows that the 
same operators are naturally represented in the 
separable Hilbert space 
${\cal H}_{\rm diff}$.

\section{Summary and discussion}

In this article I studied the loop quantization of 
diffeomorphism invariant theories of connections. 
Such theories include general relativity described in terms of 
Ashtekar-Barbero variables and extension to 
Yang-Mills fields (with or without fermions 
\cite{carlo-hugo-kiril,kirill-john-fermions}) 
coupled to gravity. For the sake of concreteness the results were 
stated for the Husain-Kucha\v{r} model \cite{husain-kuchar}, which
shares 
the phase space with general relativity, but it does not have a 
Hamiltonian constraint.

Loop quantization regularizes operators using the expression of a phase 
space function in terms of ``loop variables'' (functions of the 
holonomies of the connection along the edges of graphs 
and functions of surface smeared triads) and the quantization of 
the loop variables. 
The loop variables are a family of 
covariant functions with geometric labels whose 
quantization is a family of operators with the same geometric labels 
and an extended covariance. 
Since the quantum theory is built over the 
quantization of the loop variables, 
the extended covariance becomes a feature of the whole quantum theory.

Guidance from the classical theory tells us that the operators induced
by 
$h_l(A)$ and $h_{\pi (l)}(A)$ for any smooth map $\pi$ are gauge
related. 
In the case of non-smooth maps, 
one can not say that the functions 
$h_l(A)$ and $h_l(\f ^* (A)) := h_{\f (l)}(A)$ are gauge related since 
the non-smooth map $\f$ does not define a 
canonical transformation because 
connections of the form 
$A^\prime =\f ^* (A)$ are not in the configuration 
space of the classical theory. 
However, the quantum states are functions of generalized connections 
and $A \in \bar{\cal A}$ if and only if $\f ^* (A) \in \bar{\cal A}$ 
for any map $\f \in \bar{D}$, where $\bar{D}$ is 
a completion of the diffeomorphism group. 

Just as in the case of the internal gauge group, where 
the quantum internal gauge group is $\bar{\cal G}$, 
the same equations that defined the classical gauge group in terms of 
smooth connections are used to define the the quantum gauge group 
in terms of generalized connections. 

A quantum diffeomorphism belongs to $\bar{D}$, which 
in the analytic category is the group 
of piecewise analytic diffeomorphisms.

The resulting quantum theory yields a representation of the algebra of
observables in a {\em separable} Hilbert space. Furthermore, the 
quantum theory turns out to be equivalent to a model 
for diffeomorphism invariant gauge theories which replaces 
the space manifold with a manifestly combinatorial object \cite{c}.
Loop quantization yields a quantum theory which 
is sensitive only to the combinatorial information on the space
manifold. 
Thus, it fulfills the expectations of a framework tailored to study 
generally covariant theories. 

Since the Hamiltonian constraint is a diffeomorphism covariant
function, 
it is natural for its loop regularization to be 
$\bar{D}$ covariant (and there are versions of the Hamiltonian 
constraint which are $\bar{D}$ covariant). 
Hence, the notion of space in loop 
quantum gravity is expected to remain combinatorial after the
Hamiltonian 
constraint is imposed. 
It should be noticed that the original version of 
Thiemann's Hamiltonian constraint 
uses the Ashtekar-Lewandowski volume operator which is not 
$\bar{D}$ covariant. 
However, the modification of Thiemann's Hamiltonian 
constraint using the Rovelli-Smolin volume operator is 
$\bar{D}$ covariant, and it has been shown that it enjoys 
similar properties; in particular, the algebra of the constraints is
not 
altered by using the $\bar{D}$ covariant version of the volume 
operator \cite{wrong}. 
That the properties of the $\bar{D}$ covariant 
Hamiltonian constraint 
are the same as Thieman's is not necessarily a desirable property 
\cite{wrong}. In spite of this feature, 
a combinatorial view of loop quantization does 
suggest new treatments of dynamics.

The combinatorial picture of space suggests a simple lattice-like 
regularization of the Hamiltonian constraint. As in regular 
lattice gauge theories one can prove that the algebra of the 
constraints resembles the continuum algebra, but it has corrections 
that vanish in the continuum limit of regular lattice gauge theories. 
However, in loop quantization the continuum limit (where the 
lattice spacing, measured in a background metric, is reduced to zero) 
was replaced by the projective limit, and the correction terms do not 
vanish in the projective limit. 

I believe that there is a more promising avenue to understand 
the dynamics of loop quantum gravity. 
One can take advantage of the combinatorial 
formulation to make contact with the state sum models that arose 
borrowing ideas from topological field theories \cite{4d}. 
All the models that have been proposed up to today use the
combinatorial 
setting (or the piecewise linear setting) from the out set.

Apart from the analytic category, which I have used throughout this 
article, there is the smooth ($C^{\infty}$) category 
\cite{smooth}. The difference 
is that the allowed graphs have smooth edges; because of this, it is 
necessary to include ``wild graphs,'' which are graphs whose edges
intersect 
infinitely many times between vertices. Some aspects of this framework 
require a more careful analysis, but the quantization strategy is 
essentially the same. 
However, in view of the results of this article, 
part of the motivation to 
develop a refined version of the smooth category is lost. 
The quantum gauge group constructed by loop 
quantization is an appropriate completion of the diffeomorphism group, 
not the diffeomorphism group itself. 
Smoothness is considered as a semiclassical/macroscopic property of
space by most approaches to quantum gravity. How to reconcile this
notion with the quantization of the classical theory is a puzzling
problem. This is part of the motivation behind a proposal by Louko and
Sorkin of considering more general groups than the diffeomorphism group
as the gauge group of general relativity \cite{geroch-sorkin}.

If smoothness is not considered as fundamental, one has to find the 
characteristics of the arena of the fundamental theory. 
By completing the diffeomorphism group, loop quantization 
gives a precise replacement of classical smooth space: 
only the combinatorial information of the manifold is relevant in 
the quantum theory.

I need to acknowledge the 
illuminating conversations, suggestions and encouragement from 
Abhay Ashtekar, Alejandro Corichi, Seth Major, Roberto De Pietri, 
Jorge Pullin, Michael Reisenberger, Carlo Rovelli, Lee Smolin and 
Madhavan Varadarajan. 
Support was provided by Universidad Nacional 
Aut\'onoma de M\'exico (DGAPA), 
and grants NSF-PHY-9423950, NSF-PHY-9396246, 
research funds of the Pennsylvania 
State University, the Eberly Family research fund at PSU and the Alfred
P. 
Sloan foundation.

\end{document}